\documentclass[12pt]{article}
\usepackage{graphicx}
\usepackage{hyperref}
\newcommand{\be}{\begin{equation}}
\newcommand{\dd}{\displaystyle}
\newcommand{\ee}{\end{equation}}
\newcommand{\bea}{\begin{eqnarray}}
\newcommand{\eea}{\end{eqnarray}}

\begin{document}
\hfill{\bf BARI-TH 458/03}\par \hfill{\bf CERN-TH/2003-024}\par
\hfill{\bf UGVA-DPT-2003/02/1105}
\begin{center}
{\Large{\bf Aspects  of  the Color Flavor Locking phase of QCD in
the Nambu-Jona Lasinio approximation}}
\end{center}
\begin{center}
{\Large\bf\boldmath {}} \rm \vskip1pc {\large R.
Casalbuoni$^{a}$\footnote{On leave from the Dipartimento di
Fisica, Universita' di  Firenze, I-50019 Firenze, Italia}, R.
Gatto$^b$,
  G. Nardulli$^{c,d}$ and M. Ruggieri$^{a,c,d}$}\\ \vspace{5mm} {\it{
$^a$TH-Division, CERN, CH-1211 Geneva 23, Switzerland
 \\
 $^b$D\'epart. de Physique Th\'eorique, Universit\'e de Gen\`eve,
 CH-1211 Gen\`eve 4, Suisse\\
  $^c$Dipartimento di Fisica, Universit\`a di Bari, I-70124 Bari, Italia
  \\$^d$I.N.F.N.,
 Sezione di Bari, I-70124 Bari, Italia\\
  }}
 \end{center}

\begin{abstract}\noindent
We study two aspects of the CFL phase of QCD in the NJL
approximation. The first one is the issue of the dependence on
$\mu$ of the ultraviolet cutoff in the gap equation, which is
solved allowing a running coupling constant. The second one is the
dependence of the gap on the strange quark mass; using the high
density effective theory we perform an expansion in the parameter
$(m_s/\mu)^2$ after checking that its numerical validity is very
good already at first order.
\end{abstract}
\section{Introduction}
The existence of color superconductivity at very large densities
and low temperature is an established consequence of QCD (for
general reviews see \cite{Rajagopal:2000wf}, and
\cite{Alford:2001ab}). Since at lower densities one cannot employ
perturbative QCD, a popular approach in this regime is the use of
a Nambu-Jona Lasinio (NJL) model.  The mathematical procedure
consists in solving the mean field gap equation and selecting the
solution which minimizes the free energy. At sufficiently high
densities, for three massless flavors, the condensation pattern
leads to a conserved diagonal subgroup of color plus flavor
(color-flavor locked phase, briefly CFL phase
\cite{Alford:1997zt,Rapp:1997zu}); this phase continues to exist
when the strange quark mass is non vanishing, but not too large
\cite{Alford:1999pa,Schafer:1999pb}. An interesting property of
the CFL phase,  proved by Rajagopal and Wilczek
\cite{Rajagopal:2000ff}, is its electrical neutrality, which
implies that the densities of $u$, $d$ and $s$ quarks are equal. A
stable bulk requires not only e.m. neutrality, but also color
neutrality. It should also be a color singlet, but according to
ref. \cite{Amore:2001uf} in a color neutral macroscopic system
imposing this property does not essentially change the free
energy. Alford and Rajagopal \cite{Alford:2002kj} have shown that
in neutral CFL (with finite strange quark  mass) quarks pair with
a unique common Fermi momentum. The neutrality result is basic to
our calculations below. It allows  the use of a well defined
approximation of QCD at high density (high density effective
theory HDET), see
\cite{Hong:1998tn,Beane:2000ms,Casalbuoni:2000na} and, for a
review, \cite{Nardulli:2002ma}.

In this letter we address two aspects of CFL for QCD modeled by a
NJL four fermion interaction. Even though NJL is only a model, it
offers simple expressions that can be helpful in clarifying
physical issues; therefore a better understanding of its dynamics
is significant. The two aspects concern the role of the
ultraviolet cutoff in the NJL interaction and the relevance of the
effects due to the strange quark mass. As for the first point, the
cutoff is usually fixed once for all and considered among the
parameters of the model. However, when working  at varying large
densities the choice of the appropriate cutoff  is rather subtle.
It is suggestive to consider, for instance,  a solid where  there
is naturally a maximum frequency, the Debye frequency.  One
expects that when particles become closer the ultraviolet cutoff
extends to larger momenta. The appropriate physical simulation of
the real situation would then require a cutoff increasing with the
chemical potential, rather than a unique fixed cutoff. Our
analysis suggests that, in order to get sensible results from the
NJL model, the ultraviolet cutoff should similarly increase with
density. The results we obtain for the physical quantities, below
in this paper, show unequivocally that this is indeed the case. In
the text we will explain in more detail the essential difficulties
(such as a decreasing gap for larger densities) in which the
theory would run into if taken with a fixed cutoff. We shall
propose and apply a convenient procedure to solve the problem in
terms of a redefinition of the NJL coupling constant such as to
make it  cutoff dependent. In Section 2 we discuss this issue in
the CFL model with massless quarks and we find the optimal choice
for the dependence of the ultraviolet cutoff on the quark chemical
potential.

The second aspect we want to discuss is the role played by the non
vanishing strange quark mass. We address it in Section 3, where we
provide semi-analytical results for the dependence of the various
gap parameters on $m_s$. CFL with a massive strange quark
represents a more realistic case in which our previous discussion
can be applied; we include  both the triplet and the sextet gaps
and perform a perturbative expansion in $m_s^2/\mu^2$, obtaining
simple expressions for the first non trivial term in the
expansion. We compare this expansion with the numerical results
from the complete gap equations and we find that indeed the first
perturbative term adequately describes the full dependence. We
implement from the very beginning the electrical neutrality for
the CFL phase and, as already observed, this allows the use of the
HDET already applied for the 2SC case \cite{Casalbuoni:2002st};
for completeness also the simpler massless case is treated by the
same formalism. We notice that the gap equation with finite
strange quark mass has already been discussed in
\cite{Alford:1999pa,Alford:1999xc}. Besides the use of HDET, our
main contribution in this context is to provide semi-analytical
results for the mass dependence of the CFL gaps. We conclude the
paper with an  Appendix where we list some results and integrals
related to the gap equation.

\section{NJL running coupling constant\label{NJL}}

When the NJL interaction is used for modelling QCD at vanishing
temperature and density, one can fix  the UV cutoff $\Lambda$ such
as to get realistic quark constituent masses. Typically the cutoff
is chosen between 600 and 1000 MeV for masses ranging between 200
and 400 MeV. In any case $\Lambda$ is thought of as fixed once for
all. This gives no problems at zero density, however leads to
difficulties when one tries to simulate QCD at finite chemical
potential. In fact, at finite density one takes as relevant
degrees of freedom all the fermions with momenta in a shell around
the Fermi surface. The thickness of the shell is measured by a
cutoff $\delta$, which is the cutoff for  momenta measured from
the Fermi surface. This cutoff is chosen to be much smaller than
the chemical potential $\mu$ and much larger than the gap;
$\delta$ is  related to the NJL cutoff $\Lambda$ by the relation
$\Lambda=\mu+\delta$, because $\Lambda$ is the greatest momentum
allowed by the NJL model. This relation is however problematic
when one is interested in the behavior of the theory  for varying
$\mu$. The constraint $\Lambda=\mu+\delta$ would force $\delta$ to
vanish for increasing $\mu$, starting from $\mu<\Lambda$. In turns
this gives rise to a vanishing gap. In fact, we recall from the
simplest version of the BCS theory that the gap has a typical
behavior $\Delta\approx 2\delta\exp[-2/(G\rho)]$, where $\rho$ is
the density of the states at the Fermi surface ($\rho\propto
\mu^2$) and $G$ is the NJL coupling constant. Therefore decreasing
the volume of the shell has the effect of reducing the gap, with a
quantitatively different reduction from the state density $\rho$
and  the thickness $\delta$.  The decreasing of $\Delta$ with
$\mu$  does not correspond to the asymptotic ($\mu\to\infty$) QCD
behavior, which is characterized by an increase of the gap  with
$\mu$, though with a vanishing ratio $\Delta/\mu$
\cite{Son:1998uk}. The uncorrect behavior of $\Delta$ arises
because the model is taken to be valid only for momenta up to
$\Lambda$ which forbids  to go to values of $\mu$ of the order or
higher than $\Lambda$. Clearly this constitutes an obstacle in
physical situations where the typical chemical potential is about
400 or 500 MeV (e.g. in compact stellar objects) with a $\delta$
of the order 150 or 200 MeV. In fact it turns out difficult, if
not impossible, to explore higher values of $\mu$ for any
reasonable choice of $\Lambda$.

In this paper we make a proposal to overcome this situation. We
start by noticing that the phenomenology at zero temperature and
density\footnote{For the sake of discussion we consider here the
ideal case of massless quarks, the more realistic case of a
massive strange quark will be considered in the following} is
completely determined by the meson decay coupling constant $f_\pi$
and by the constituent quark mass, or equivalently by the chiral
condensate. In the NJL model these two quantities are fixed by two
equations (see for a review \cite{Klevansky:qe}), one fixing
$f_\pi$ and the other the chiral gap equation. These equations
depend on the cutoff $\Lambda$ and on the NJL coupling $G$. Our
proposal consists in assuming the coupling $G$ to be a function of
$\Lambda$ in such a way that $f_\pi$ assumes its experimental
value and that the chiral gap equation is satisfied for any choice
of $\Lambda$.

To be more explicit, we  write the  Nambu-Jona Lasinio equations
with a three dimensional cutoff $\Lambda$ \cite{Klevansky:qe}; the
equation for the $\pi$ leptonic decay constant reads : \be
f_{\pi}^2=3m^{*\,2}
\int\frac{d^3p}{(2\pi)^3}\frac{1}{E_p^3}\,\theta\left({\Lambda-|\vec
p |}\right)=-\frac{3m^{*\,2}}{2\pi^2}
\left[\frac{\Lambda}{\sqrt{m^{*\,2}+\Lambda^2}}-{\rm arcsinh}
\frac{\Lambda}{m^{*}}\right]\ ,\label{fpi}\ee where $m^*$ is the
constituent mass at $\mu=0$ which is determined by the
self-consistency condition: \be m^*=m_0+\frac{4m^*
}{3\pi^2}G(\Lambda)\int_0^{\Lambda}\frac{p^2dp}{\sqrt{p^2+m^{*\,^2}}}\
;\label{self} \ee $m_0$ is the quark current mass which is assumed
in this Section to be zero. $G(\Lambda)$ is the NJL coupling
having dimension mass$^{-2}$; it could be understood as the effect
of a fictitious gluon propagator \cite{Alford:1997zt}:\be
i\,D^{\mu\nu}_{ab}=\,i\,\frac{g^{\mu\nu}\delta^{ab}}{\Lambda^2} \
,\ee and $\Lambda^2G(\Lambda)$ would take the role of the square
of the strong coupling constant. Eq. (\ref{fpi}), with $f_\pi=93$
MeV, implicitly defines the function $m^*=m^*(\Lambda)$ which we
use in eq. (\ref{self}) to get the function $G=G(\Lambda)$. The
result of this analysis is in fig. 1.
\begin{figure}[h]
\begin{center}
\includegraphics[width=8cm]{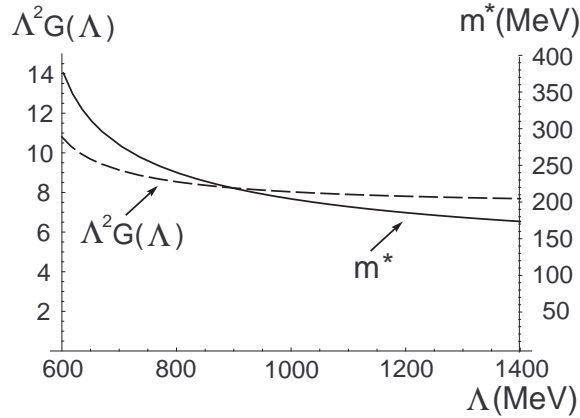}
\end{center}
\caption{\footnotesize The running NJL coupling constant $G$
(dashed line) and the running constituent mass $m^*$ (in MeV,
solid line) as functions of $\Lambda$ (in MeV). $\Lambda$ is the
ultraviolet cutoff. The vertical axis on the left refers to
$\Lambda^2G(\Lambda)$, while the axis on the right refers to
$m^*$. \label{l}}
\end{figure}

\begin{figure}[h]
\begin{center}
\includegraphics[width=8cm]{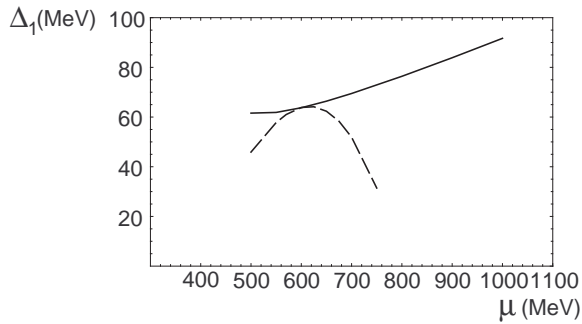}
\end{center}
\caption{\footnotesize The CFL gap $\Delta$ for massless quarks,
as obtained from eq. \ref{2.186}, versus the quark chemical
potential for the two cases discussed in the text. Solid line:
running NJL coupling $G(\mu+\delta)$ and cutoff $\delta=c\mu$,
with $c=0.35$; dashed line: $\delta=\Lambda-\mu$, where
$\Lambda=800$ MeV, and $G(\Lambda)=13.3$ GeV$^{-2}$. The picture
shows the different qualitative behavior with $\mu$ of the gap.
\label{2}}
\end{figure}

Our choice implies  that  a NJL model is defined at any scale by
using the appropriate $G(\Lambda)$. Whereas in the usual case we
have to keep the momenta smaller than the cutoff, now, for any
given momenta, we can fix the cutoff in such a way that it is much
bigger than the  momenta. The phenomenology of the chiral world is
clearly unaffected by this procedure; in fact it turns out that
the constituent mass acquires a weak dependence on the cutoff, and
therefore it can be fixed at the most convenient value. Also the
quantity $G(\Lambda)\Lambda^2$ decreases weakly with the cutoff.
In applying these considerations to the calculations at finite
density, we have only to use the appropriate value of the coupling
as given by $G(\mu+\delta)$, where now $\mu+\delta$ has nothing to
do with the value of $\Lambda$ chosen to fit the chiral world. To
give an explicit example we consider  the CFL phase with massless
quarks.
 There are  two independent gaps $\Delta$ and $\Delta_9$
($\Delta_9=-2\Delta$ if the pairing is only in the antitriplet
channel) and the gap equations are
\cite{Alford:1997zt,Nardulli:2002ma}: \bea \Delta&=&
-\,\frac{\mu^2G}{6\pi^2} \left(\Delta_9 \ {\rm arcsinh}\frac\delta
{|\Delta_9|}-2\Delta\ {\rm arcsinh}\frac\delta {|\Delta|}
\right)\, ,\cr
 \Delta_9&=& -\,\frac{4\mu^2 G\Delta}{3\pi^2}\
 {\rm arcsinh}\frac\delta
{|\Delta|}\ .\label{2.186}\eea If one uses a fixed value for
$\Lambda=\mu+\delta$, as for instance in ref.
\cite{Steiner:2002gx}, one gets a non monotonic behavior of the
gap, as it can be seen from fig. \ref{2}, (dashed line); a similar
behavior was found in \cite{Steiner:2002gx} (their fig. 1), albeit
a different choice of the parameters produces some numerical
differences.
On the other hand the solid line shows an increasing behavior of
the gap. We see that in this way one reproduces
 qualitatively the behavior found in QCD for asymptotic chemical
 potential \cite{Son:1998uk}.
This result is obtained by the running NJL coupling
$G(\mu+\delta)$, with the following choice of the cutoff $\delta$:
\be \delta=c\,\mu\label{delta}\ee with
 $c$ a fixed constant ($c=0.35$ in fig. \ref{2}). The reason for this choice is that,
as discussed above,  when increasing $\mu$, we do not want to
reduce the ratio of the number of the
  relevant degrees of freedom  to the volume of the Fermi sphere.
Requiring the fractional importance to be constant is equivalent
to require eq. (\ref{delta}).

\begin{figure}[h]
\begin{center}
\includegraphics[width=10cm]{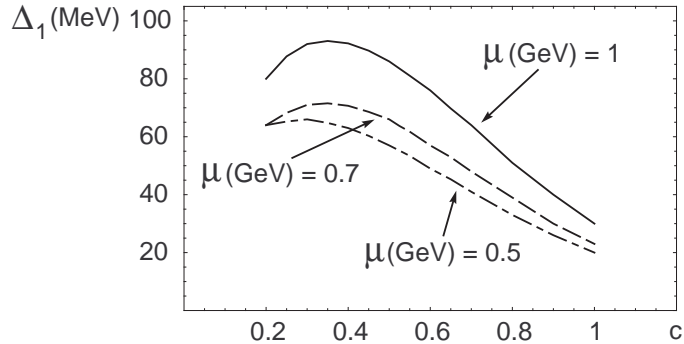}
\end{center}
\caption{\footnotesize Gap parameter $\Delta_{1}$ as a function of
$c$, where $\delta=c\,\mu$; the quarks are massless. The three
curves are obtained for three values of the quark chemical
potential: $\mu=1.0,\,0.7,\,0.5$ GeV.} \label{GapDeltaMassless}
\end{figure}

In fig. \ref{GapDeltaMassless}, we plot the gap parameter
$\Delta_1$  (the results for $\Delta_9$ are similar) as a function
of $c$ for three different values of the chemical potential. In
general there exists a window of values for $c$: \be
c=0.35\pm0.10\ ,\ee  where the gap parameters are less dependent
on $c$. This is therefore the range of $c$ we shall assume  below.
It can be also noted that the  result for the QCD superconducting
model we are considering here has its
 counterpart in  solid state physics, where the analysis of \cite{russi} shows
a linear increase of $\delta$ with $\mu$ similar to eq.
(\ref{delta}).

 These results, valid for the massless case, are confirmed by a
 complete numerical analysis including the effect of a strange mass.
This will be discussed in the subsequent section.
\section{Effective lagrangian
for gapless quarks}

The massive case is considerably more involved, and the simple set
of equations (\ref{2.186}) has to be substituted by a system of 5
equations \cite{Alford:1999pa} that can be only solved
numerically. In \cite{Steiner:2002gx} the CFL phase with massive
strange quark was also considered. In comparison with
\cite{Alford:1999pa} the derivation we present here has the
advantage of offering semi-analytical results, thanks to an
expansion in powers of $m_s/\mu$. We principally differ from  ref.
\cite{Steiner:2002gx} for the different treatment of the cutoff,
as discussed in the previous section, and for the inclusion of
pairing in both the antitriplet and the sextet color channel. The
possibility of a semi-analytical treatment rests on the HDET
approximation.  This effective lagrangian approach was extended in
\cite{Casalbuoni:2002st} to the 2SC phase with massive quarks and
here we treat  the three flavor case.

In the HDET one introduces effective velocity dependent fields, corresponding
to the positive energy solutions of the field equations:
$
  \psi_{\alpha, i,\vec n}(x)$ where $\alpha,\, i$ are color and flavor indices,
  $\vec n=\vec v/|\vec
v|$, and $\vec v$ is the quark velocity defined by
 the equation \be p^\mu=\mu v^\mu+\ell^\mu \label{dec}\ee with $v^\mu=(0,\vec v)$.
The effective fields $\psi_{\alpha, i\,\vec n}(x)$ are expressed in terms of
Fourier-transformed quark fields $\tilde\psi_{\alpha, i}(p)$
as follows
 \be
\psi_{\alpha, i\,\vec n}(x)={\cal P}_{+}
\int\frac{d^{4}p}{(2\,\pi)^{4}}\,e^{i\,(p-\mu v)\cdot
x}\,\tilde\psi_{\alpha, i}(p) \ .\ee
Here ${\cal P}_{+}$ is
the positive energy projector, defined, together with ${\cal P}_{-}$,
by the formula
\begin{equation}
{\cal
P}_{\pm}=\frac{1\pm(\vec{\alpha}\cdot\vec{v}+x_i\,\gamma^{0})}{2}
\label{energyprojectors1} \ ,
\end{equation} where $ x_i=m_i/\mu$ and $m_i$ is the mass of the quark having flavor $i$.
We now change the color-flavor basis introducing new fields
 $\psi^A_{\vec n}$ as follows:
\begin{equation}
\psi_{\alpha, i,\vec n}(x)\,=\,\sum_{A=1}^{9} \frac{\lambda^A_{\alpha i}}
{\sqrt 2} \, \psi^A_{\vec n}(x)\ ,
\end{equation}
where $\lambda_A$ with $A=1,\dots,8$ are the usual Gell-mann
matrices and $\lambda_9=\sqrt{2/3}\,\lambda_0$. We also introduce
 the Nambu-Gor'kov
doublet
\begin{equation}
\chi_A=\frac{1}{\sqrt{2}}\left(\begin{array}{c}
\psi^A_{\vec{n}}\\
C\,\psi^{A,*}_{-\vec{n}}
\end{array}\right)\ .
\end{equation}
In the $\chi_A$ basis the lagrangian of the quarks, including the
quark-gluon interaction and the gap therm can be written in
momentum space as follows: \be{\cal L}={\cal L}_0+{\cal L}_1+{\cal
L}_\Delta\ .\ee ${\cal L}_0$ is the kinetic term while ${\cal
L}_1$ describes the quark-gluon interaction. They are given by:
\begin{eqnarray}
{\cal L}_0&=&\sum_{\vec{n}}\sum_{A,B=1}^9\,\chi_{A}^{\dagger}\,
\left(\begin{array}{cc}
{\bf \,V_{AB}\cdot \ell}&0\\
0&{\bf \,\tilde{V}_{AB}\cdot \ell}
\end{array}\right)\,\chi_B\ ,\cr
{\cal L}_1
&=&\,ig\,\sum_{\vec{n}}\sum_{A,B=1}^9\,\chi_{A}^{\dagger}\,
\left(\begin{array}{cc}
{\bf H_{AB}\cdot}A&0\\
0&-{\bf \tilde{H}_{AB}}^*\cdot A
\end{array}\right)\,\chi_B\ .
\label{kineticRotatedTWO}
\end{eqnarray}
We have introduced the symbols \bea {\bf \,V_{AB}\cdot \ell}&=&
Tr\left[T_A\,T_B\,{\cal V}^\mu\right] \ell_\mu~,~~~~~~~~~~~~~  {\bf\tilde
V_{AB}\cdot \ell}= Tr\left[T_A\,T_B\,\tilde {\cal V}^\mu\right]\ell_\mu
\label{VmatrixDef} \cr {\bf \,H_{AB}\cdot }A&=&H^\mu_{AmB}\cdot A^m_\mu=
\,\frac{1}{\sqrt 2}\,Tr\left[T_A\,T_m\,T_B\,{\cal V}^\mu\right]A^m_\mu ~,\cr
{\bf\tilde H_{AB}\cdot  }A&=&{\tilde H}^\mu_{AmB}\cdot A^m_\mu=
\,\frac{1}{\sqrt 2}\,Tr\left[T_A\,T_m\,T_B\,\tilde {\cal V}^\mu\right]A^m_\mu
\label{Hmatrix}\ . \eea In these equations   $T_A=\lambda_A/\sqrt{2}$,
$A^m_\mu$ is the gluon field, and ${\cal V}^\mu$ denotes the matrix \be {\cal
V}^\mu_{ij}= \left(\begin{array}{ccc}
{\cal V}^\mu_u&0&0\\
0&{\cal V}^\mu_d&0\\0&0&{\cal V}^\mu_s
\end{array}\right)\ee with
${\cal V}^\mu_i =(1,\,\vec v_i)$ for each flavor $i$; a similar
definition holds for $\tilde{\cal V}^\mu_{ij}$, with $\tilde{\cal
V}^\mu_i =(1,\,-\,\vec v_i)$. In the limit $m_u=m_d\approx 0$ one
has $ v_u=v_d=1\ ,~~~v_s=\sqrt {1-x_s^2}=\sqrt {1-m_s^2/\mu_s^2}$.

\begin{figure}[th]
\begin{center}
\includegraphics[width=8cm]{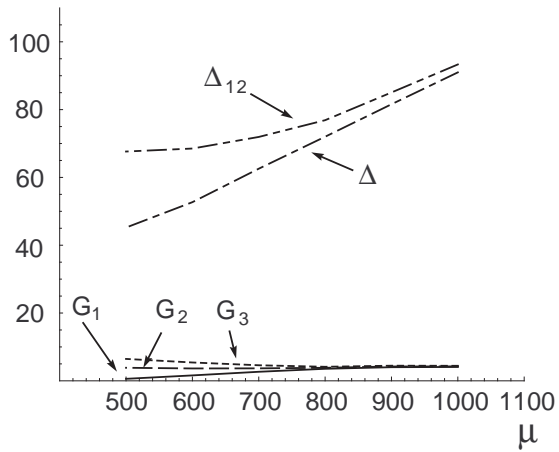}
\end{center}
\caption{\footnotesize The five gap parameters vs. the chemical
potential $\mu$; results obtained by the numerical solution of the
exact gap equations, for $m_s=250$ MeV and $\delta=0.35\mu$. The
upper curve refers to $\Delta_{12}$, while the middle one refers
to $\Delta$. The lower curves are the data obtained for the three
sextet gap parameters $|G_1|$, $|G_2|$ and $|G_3|$. Gaps and $\mu$
are expressed in MeV.} \label{PlotLavoro1}
\end{figure}

Let us now turn to the gap term ${\cal L}_\Delta$. We consider CFL
condensation in both the antisymmetric $\bar{\bf 3}_A$ and in the
symmetric ${\bf 6}_S$ channels. We assume equal masses (actually
zero) for the up and down quarks and neglect quark-antiquark
chiral condensates, whose contribution is expected to negligible
in the very large $\mu$ limit. Also the contribution from the
repulsive ${\bf 6}_S$ channel is expected to be small, but we
include it because the gap equations
 are consistent only with condensation in both the ${\bf 6}_S$ and the $\bar{\bf
3}_A$ channels.
The condensate we consider is therefore
\begin{equation}
<\psi_{\alpha i}\,C\,\gamma_5\,\,\psi_{\beta j}>\sim
\left(\Delta_{ij}\,\epsilon^{\alpha\beta I}\epsilon_{ijI}
+G_{ij}\,(\delta^{\alpha i}\,\delta^{\beta j}+\delta^{\alpha
j}\,\delta^{\beta i})
\right)\ .
\label{6plus3condensate}
\end{equation}
The first term on the r.h.s
accounts
 for the condensation in the  $\bar{\bf
3}_A$  channel and the second one describes condensation in ${\bf
6}_S$ channel. As we assume $m_u=m_d$, we put \be
\Delta_{us}=\Delta_{ds}\equiv\Delta\
,~~~~~~~~~~~~\Delta_{ud}\equiv \Delta_{12}\ ,\ee \be
G_{uu}=G_{du}=G_{ud}=G_{dd}\equiv G_1\
,~~~~~~~~~~~~G_{us}=G_{ds}\equiv G_2\ ,~~~~~~~~~~~ G_{ss}\equiv
G_3 \ ,\ee which reduces the number of independent gap parameters
to five. We stress that we impose electrical and color neutrality,
which, as shown  in ref. \cite{Rajagopal:2000ff},  is indeed
satisfied in the color-flavor locked phase of QCD because in this
phase the three light quarks number densities are equal $
 n_u=n_d=n_s$, with no need for electrons, i.e. $n_e=0$. As a consequence,
the  Fermi momenta of the three quarks are equal:
\be
p_{F,\,u}\,=\,p_{F,\,d}\,=\,p_{F,\,s}\,\equiv\,p_F\,
\ ,\label{p}\ee
which, in terms of the quark chemical potentials and Fermi velocities
 can be written as
\be\mu_u|\vec v_u|\,=\,\mu_d|\vec v_d|\,=\,\mu_s|\vec v_s|\ .\ee
It follows that  the wave function of the quark-quark condensate
has no dependence on the Fermi energies and the condensation can
be described in the mean field approximation
 by the following lagrangian term containing only the effective fields $\psi_{\alpha i,\,\pm}$:
\begin{equation}
{\cal L}_\Delta=-\frac{\Delta_{ij}\,\epsilon^{\alpha\beta I}\epsilon_{ijI}+
G_{ij}\,(\delta^{\alpha i}\,\delta^{\beta j}+\delta^{\alpha
j}\,\delta^{\beta i})}{2}\,
\sum_{\vec{n}}\,\psi^T_{\alpha i,\,-}\,C\,\gamma_5\,\psi_{\beta j,\,+}\,+\,h.c.\label{Ldelta}
\end{equation}
Using the Nambu-Gor'kov fields one rewrites (\ref{Ldelta}) as follows:
\be{\cal L}_\Delta=
\sum_{\vec{n}}\sum_{A,B=1}^9\,\chi_{A}^{\dagger}\,
\left(\begin{array}{cc}
0&\gamma_5\,\otimes \,{\bf \Delta_{AB}}\\
\gamma_5\,\otimes \,{\bf \Delta_{AB}^\dagger}&0
\end{array}\right)\,\chi_B
\label{kineticRotatedTWO}
\ee
where\be{\bf \Delta_{AB}}={\bf\Delta_{AB}}({\bar{\bf
3}})+{\bf\Delta_{AB}}({\bf
6})\ee and
\begin{equation}
{\bf\Delta_{AB}}({\bar{\bf 3}})=\left(
\begin{array}{cccc}
\Delta_{12} \,I_{3}&0&0&0\\
0&\Delta\, I_4&0&0\\
0&0&\frac{1}{3}\left(4\,\Delta-\Delta_{12}\right)&\frac{\sqrt{2}}{3}\left(\Delta
-\Delta_{12}\right)\\
0&0&\frac{\sqrt{2}}{3}\left(\Delta-\Delta_{12}\right)&-\frac{2}{3}\left(\Delta_{
12}+2\,\Delta\right)
\end{array}
\right)
\end{equation}\begin{equation}
{\bf \Delta_{AB}}({\bf 6})=\left(
\begin{array}{cccc}
G_1\,I_3&0&0&0\\
0&G_2\,I_4&0&0\\
0&0&G_1-\frac{4}{3}(G_2-G_3)&\frac{\sqrt{2}}{3}(3G_1-2G_3-G_2)\\
0&0&\frac{\sqrt{2}}{3}(3G_1-2G_3-G_2)&\frac{2}{3}(3G_1+2G_2+G_3)
\end{array}\right)\ ,
\label{symmetricGapMatrix}\\
\end{equation} where
$I_n$ stays for the $n\times n$ identity matrix. The nine
eigenvalues of the matrix $ {\bf \Delta_{AB}}$ are reported in the
Appendix. From the lagrangian \be {\cal L}_0\,+\, {\cal L}_\Delta=
\sum_{\vec{n}}\sum_{A,B=1}^9\,\chi_{A}^{\dagger}\,
\left(\begin{array}{cc}
{\bf \,V_{AB}\cdot \ell}&\gamma_5\,\otimes \,{\bf \Delta_{AB}}\\
\gamma_5\,\otimes \,{\bf \Delta_{AB}^\dagger}&{\bf \,\tilde{V}_{AB}\cdot \ell}
\end{array}\right)\,\chi_B\,=\sum_{\vec{n}}\sum_{A,B=1}^9\,\chi_{A}^{\dagger}S^{-1}_{AB}(\ell)
\chi_B\ee the fermionic propagator is  obtained:
\begin{displaymath}
S_{AB}(\ell)=\left(\begin{array}{cc}
{\bf\frac{\dd 1} {\dd \Delta\tilde{V}\cdot  \ell\Delta^{-1}\,V\cdot
l-\Delta^2}\,\Delta\tilde{V}\cdot \ell\Delta^{-1}}
&{\bf -\frac{\dd 1}{\dd \Delta\,\tilde{V}\cdot  \ell\,\Delta^{-1}\,V\cdot
l-\Delta^2}\,\gamma_5\,\otimes \,\Delta}\\ &\\
{\bf -\frac {\dd 1} {\dd \Delta V\cdot \ell\,\Delta^{-1}\,\tilde{V}\cdot
l-\Delta^2}\,\gamma_5\,\otimes \Delta}&
{\bf \frac{\dd 1} {\dd \Delta V\cdot  \ell\,\Delta^{-1}\tilde{V}\cdot
l-\Delta^2}\,\Delta V\cdot  \ell\,\Delta^{-1}}
\end{array}\right)_{AB}
\label{fermionicFULL}
\end{displaymath}
The off-diagonal matrix $S_{AB}^{12}$ (1, 2= NG indices) is the
anomalous component of the quark propagator, which is what we need
to write down the gap equations. In matrix form they are as
follows
\begin{equation}
-{\bf
\Delta_{AB}}=-\frac{i\,G(\mu+\delta)\,p_F^2}{4\,\pi^3}\,H_{AaC}^\mu\,H_{DbB}^{\nu
*}\,\int_{-\delta}^{+\delta}
 d\ell_\parallel\int_{-\infty}^{+\infty}d\ell_0\delta^{ab}\,g_{\mu\nu}\,S^{12}_{CD}(l)\,,
\label{FullGapEquation}
\end{equation}
where the color-flavor symbols $H_{AaB}^\mu$ have been defined in
(\ref{Hmatrix}) and the running NJL coupling is computed at
$\mu+\delta$, according to the previous discussion. In eq.
(\ref{FullGapEquation})  $\delta=c\mu$ is the  cutoff discussed in
section \ref{NJL}. The gap equations can be numerically solved;
results  for $m_s=250$ MeV and $c=0.35$ are reported in Fig.
\ref{PlotLavoro1}.

A semi-analytical solution can be found by performing an expansion
in the  strange quark  mass: \be x_s=\frac{m_s}{\mu_s}
\approx\frac{m_s}{\mu}\ll 1 \ . \ee Here $\mu=\mu_u=\mu_d$,
$\mu_s=\mu+\delta\mu$ with $\delta\mu={\cal O}(m_s/\mu)^2$. We
define:\bea \Delta_1&=&\Delta_{12}+G_1\equiv\Delta^0_1+\epsilon_1
x_s^2~,~~~~
\Delta_9=-\,2\,\Delta_{12}+4\,G_1\equiv\Delta^0_9+\epsilon_9
x_s^2\, ,\cr&&\cr
\Delta&\equiv&\frac{4\Delta_1^0-\Delta_9^0}{6}-\xi_1 x_s^2~,~~
G_2\equiv\frac{2\Delta_1^0+\Delta_9^0}{6}-\xi_2 x_s^2~,~~
G_3\equiv\frac{2\Delta_1^0+\Delta_9^0}{6}-\xi_3 x_s^2 \
,\label{full}\eea where $\Delta^0_1, \Delta^0_9$ are the values
for massless quarks and can be obtained from eqns. (\ref{2.186}).
For $m_s\neq 0$ the number of  gaps increases from 2 to 5, but
eqns. (\ref{full}) give immediately the solutions for any value of
$m_s$ (with the proviso $x_s=m_s^2/\mu^2\ll 1$) if one knows the
parameters $\xi=(\epsilon_1,\epsilon_9,\xi_1,\xi_2,\xi_3)$. They
can be obtained by solving the  system of 5 linear algebraic
equations \be {\bf A}\cdot{\bf \xi}={\bf f}\ . \ee The matrices
${\bf A}$ and $f$  are reported in the appendix. Numerical results
for the parameters $\bf \xi$ are in Table 1. For a strange mass of
$250$ MeV we find the rate $\Delta_9/\Delta_1$ equal to $-2.365$
for $\mu=500$ MeV and $-2.33$ for $\mu=1000$ MeV.
 \begin{table}[h]
\begin{center}
\begin{tabular}{|c|c|c|c|c|c|}
\hline
$\mu$&$\epsilon_1$&$\epsilon_9$&$\xi_1$&$\xi_2$&$\xi_3$\\
\hline
500&-35.29&90.48&33.63&-2.06&-6.20\\
\hline
700&-44.86&102.93&41.16&-3.10&-8.72\\
\hline
1000&-61.21&135.08&55.69&-4.42&-12.23\\
\hline
\end{tabular}
\end{center}
\caption{{\footnotesize Values of the ${\xi}$ components, obtained
by means of the $5\times5$ system of linear equations. In the gap
equations we have made the choice $\delta=0.35\,\mu$. All the gaps
and $\mu$'s are expressed in MeV.} \label{tableCSIvector}}
\end{table}

\begin{figure}[h]
\begin{center}\includegraphics[width=15cm]{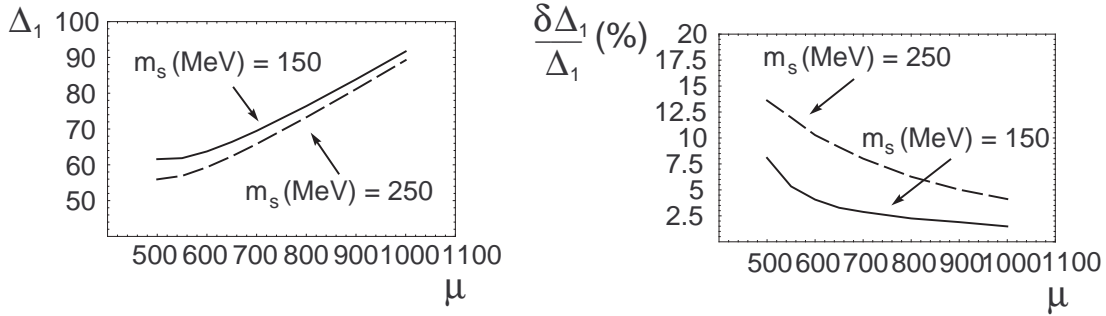}
\end{center}
\caption{\footnotesize Gap parameter $\Delta_1=\Delta_{12}+G_1$ as
a function of $\mu$. Left panel: results for two different values
of the strange quark mass, $m_s=150,\,250$ MeV (curves are
obtained by means of the perturbative gap equations). Right panel:
relative variation
$\delta\Delta_1/\Delta_1=[\Delta_1(m_s)-\Delta_1(0)]/\Delta_1(0)$.
In this plot $\delta=0.35\mu$. Gaps and $\mu$'s are expressed in
MeV.\label{fig1}}
\end{figure}
\begin{figure}[h]
\begin{center}\includegraphics[width=8cm]{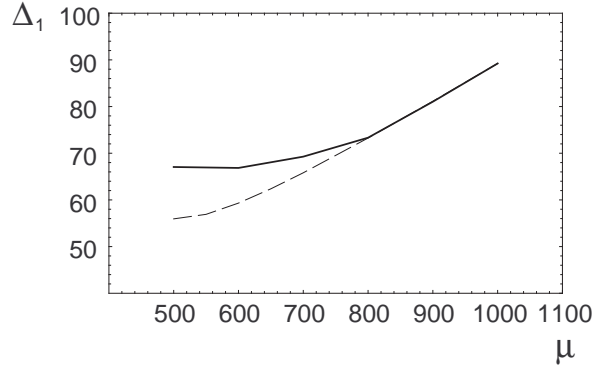}
\end{center}
\caption{\footnotesize This diagram shows the difference between
the exact numerical calculation of the gap $\Delta_1$ for
$m_s=250$~MeV (solid line)  and the approximate result obtained
through the expansion in $m_s$ (dashed line). Units are MeV.
\label{fig6}}
\end{figure}

From these results we  can compute the dependence of the various
gaps on $\mu$. For $\Delta_1$ it is reported (for two values of
the strange quark mass) in fig.\ref{fig1}.  We stress that the
approximation we consider here is indeed very good  in the range
of parameters we have considered in the present paper, i.e. $
m_s=150-250$ MeV and $\mu=500-1000$ MeV. As an example, fig.
\ref{fig6} shows the approximate solution for $\Delta_1$ (solid
line),  which differs only by  a few percent from the exact
solution (dashed line), except in the lower region of $\mu$. These
results are obtained for a mass of 250 MeV; for smaller values the
effect is even less relevant.

\section{Conclusions}

A main point in this work has been to clarify the issue of the
dependence on $\mu$ of the ultraviolet cutoff in the NJL model and
its application to CFL phase of QCD. A convenient procedure
consists in re-defining the NJL coupling such as to make it cutoff
dependent. The application  to the CFL model with $m_s\not=0$
leads, in the approximation of the high density effective theory,
to an expansion in the parameter $(m_s/\mu)^2$, whose numerical
validity is very good already at first order. Both triplet and
sextet gaps are included. The gap parameters are obtained as
functions of the chemical potential and their behavior is
consistent with the expected asymptotic limits.

\section*{Appendix} {\bf Eigenvalues of the gap matrix}

The eigenvalues of the gap matrix for CFL with massive strange
quark are as follows: $\Delta_1=\dots=\Delta_5=\Delta_{12}+G_1$,
$\Delta_6=\Delta_7=\Delta+G_2$, $
\Delta_{8/9}=\frac{1}{2}\left(x-z\pm\sqrt{(x+z)^2+4y^2}\right)$
where $x,y,z$ are given by
\begin{eqnarray}
x&=&\frac{1}{3}(4\Delta-\Delta_{12})+G_1-\frac{4}{3}(G_2-G_3)\, ,\\
y&=&\frac{\sqrt{2}}{3}(\Delta-\Delta_{12})+
\sqrt{2}\left(G_1-\frac{G_2}{3}-\frac{\sqrt{2}G_3}{3}\right)\,,\\
z&=&\frac{2}{3}(\Delta_{12}+2\Delta)-\frac{2}{3}(3G_1+2G_2+G_3)\,.
\end{eqnarray}

{\bf  Matrices ${\bf \xi}$, $ {\bf A}$, $\bf f$}

We write down explicitly the matrices ${\bf \xi}$, $ {\bf A}$,
$\bf f$ defined in the text
 \be {\bf \xi}^T=\left( \epsilon_1 \ \epsilon_9\
\xi_1\ \xi_2\ \xi_3 \right)\ ;
\end{equation}
\footnotesize
\begin{equation}{\bf A}
= \left(\begin{array}{ccccc} \alpha&\beta&-\frac{2}{9}b(-g_1+2g_9-g_{19})&
\frac{2}{9}b(-g_1+2g_9-g_{19})&\frac{2}{9}b(g_1-g_9+2g_{19})\\
\alpha&\beta&-1-\frac{b}{9}(7g_1+4g_9+g_{19})&
-1+\frac{b}{9}(g_1+4g_9+g_{19})&-\frac{b}{9}(4g_1+2g_9+2g_{19})\\
\alpha&\beta& -\frac{4}{3}-\frac{2b}{9}(5g_1+2g_9+g_{19})&
\frac{4}{3}-\frac{2b}{9}(7g_1-2g_9-g_{19})&
-\frac{4}{3}-\frac{2b}{9}(-g_1+g_9+2g_{19})\\
0&0& -\frac{\sqrt{2}}{3}\left(1+\frac{b}{3}(5g_1-g_{19})\right)&
\frac{\sqrt{2}}{3}\left(1-\frac{b}{3}(g_1-g_{19})\right)&
-\frac{2\sqrt{2}}{3}\left(1+\frac{b}{3}(g_1-g_{19})\right)\\
\gamma&1& \frac{4}{3}-\frac{2b}{9}4g_1& -\frac{4}{3}-\frac{16b}{9}g_1&
\frac{2}{3}+\frac{4b}{9}g_1
\end{array}\right)
\ee\normalsize
 where  we have defined
$b=G(\mu+\delta)\mu^2/2\pi^2$, $\alpha=1-\frac{2}{3}b\,h_1$,
$\beta=\frac{1}{3}b\,h_9$, $\gamma=\frac{8b}{3}h_1$; finally \be
{\bf f}=\left(\begin{array}{c}
b\left(-\frac{\tilde{I}^0_1}{3}+\frac{\tilde{I}^0_9}{6}
-\frac{1}{18}(f_1+f_9-4f_{19})\right)\cr
b\left(-\frac{\tilde{I}^0_1}{2}+\frac{\tilde{I}^0_9}{4}-
\frac{1}{18}(5f_1-f_9-2f_{19})\right)\cr
b\left(\frac{5}{18}\left(-2\tilde{I}^0_1+2\tilde{I}^0_9\right)
+\frac{1}{18}(11f_1-f_9-4f_{19})\right)\\
b\left(\frac{1}{18\sqrt{2}}(-7\tilde{I}^0_1-\tilde{I}^0_9)+
\frac{\sqrt{2}}{18}(\frac{17}{2}f_1+f_{19}))\right)\cr
b\left(\frac{16}{9}\tilde{I}^0_1+\frac{10}{18}f_1\right)
\end{array}\right)\ .\ee

 The values of the parameters $f_j$, $g_j$, $\tilde{I}^0_j$, $h_k$
are listed below. \bea f_1&=&\frac{1}{2\pi i}\Delta^0_1
J^\delta(\Delta^0_1)~,~~~ f_9=\frac{1}{2\pi i}\Delta^0_9
J^\delta(\Delta^0_9)~,~~~ f_{19}=\frac{1}{2\pi i}\Delta^0_1
J^\delta(\Delta^0_1,\Delta^0_9)\ , \cr&&\cr g_1&=&\frac{1}{2\pi
i}I^\delta(\Delta^0_1)~,~~~ g_9=\frac{1}{2\pi i} I^\delta(\Delta^0_9)~,~~~
g_{19}=\frac{1}{2\pi i} I^\delta(\Delta^0_1,\Delta^0_9)\ ,\\
\tilde{I}^0_j&=&\Delta^0_j {\rm
arcsinh}\left|\frac{\delta}{\Delta^0_j}\right| ~,~~~h_k={\rm
arcsinh}\left|\frac{\delta}{\Delta^0_j}\right|-1 \ .\eea In this
expression we use the following parametric
 integrals:
\begin{eqnarray}
I^\delta(x)&=&\int_{-\delta}^\delta d^2 l\,\frac{V\cdot l\,\tilde{V}\cdot
l+x^2}{D(x)^2} =2\,\pi\,i\left(-{\rm
arcsinh}\left|\frac{\delta}{x}\right|+\frac{\delta}{\sqrt{\delta
^2+x^2}}\right)\ ,
\\
I^\delta(x_1,x_2)&=&\int_{-\delta}^\delta d^2 l\,\frac{V\cdot l\,\tilde{V}\cdot
l+x_1\,x_2}{D(x_1)\,D(x_2)}=
\nonumber\\
&=&\frac{2\,\pi}{i}\left[{\rm arcsinh}\left|\frac{\delta}{x_2}\right|
+\frac{x_1(x_1+x_2)}{x_2^2-x_1^2}\left({\rm
arcsinh}\left|\frac{\delta}{x_2}\right|-{\rm
arcsinh}\left|\frac{\delta}{x_1}\right|\right)
\right]\ ,\\
J^\delta(x)&=&\int_{-\delta}^\delta d^2 l\,\frac{l_\parallel^2}{D(x)^2}
=i\,\pi\left({\rm
arcsinh}\left|\frac{\delta}{x}\right|-\frac{\delta}{\sqrt{\delta^2+x
^2}}\right)\,,\\
J^\delta(x_1,x_9)&=&\int_{-\delta}^\delta d^2
l\,\frac{l_\parallel^2}{D(x_1)D(x_9)} =\frac{i\,\pi}{2(x_1^2-x_9^2)}\left[
-2\,\delta\left(\sqrt{x_1^2+\delta^2}-\sqrt{x_9^2+\delta^2}\right)\right.
\nonumber\\
&&\left. +2\,x_1^2\,{\rm arcsinh}\left|\frac{\delta}{x_1}\right|
-2\,x_9^2\,{\rm arcsinh}\left|\frac{\delta}{x_9}\right| \right]\ .
\end{eqnarray}

\begin{center}
{\bf{Acknowledgements}} \end{center}

We wish to thank M. Mannarelli for useful discussions. One of us,
G. N., wishes to thank the CERN theory group for the very kind
hospitality.


\begin{thebibliography}{99}

\bibitem{Rajagopal:2000wf}
K.~Rajagopal and F.~Wilczek,
[arXiv:hep-ph/0011333].

\bibitem{Alford:2001ab}
M.~Alford, {\it Ann. Rev. Nucl. Part. Sci.} {\bf 51} (2001) 131
[arXiv:hep-ph/0102047].


\bibitem{Alford:1997zt}
M.~G.~Alford, K.~Rajagopal and F.~Wilczek,
Phys.\ Lett.\ B {\bf 422} (1998) 247 [arXiv:hep-ph/9711395].

\bibitem{Rapp:1997zu}
R.~Rapp, T.~Schaefer, E.~V.~Shuryak and M.~Velkovsky,
Phys.\ Rev.\ Lett.\  {\bf 81} (1998) 53 [arXiv:hep-ph/9711396].

\bibitem{Alford:1999pa}
M.~G.~Alford, J.~Berges and K.~Rajagopal,
Nucl.\ Phys.\ B {\bf 558} (1999) 219 [arXiv:hep-ph/9903502].

\bibitem{Schafer:1999pb}
T.~Schaefer and F.~Wilczek,
Phys.\ Rev.\ D {\bf 60} (1999) 074014 [arXiv:hep-ph/9903503].

\bibitem{Rajagopal:2000ff}
K.~Rajagopal and F.~Wilczek,
Phys.\ Rev.\ Lett.\  {\bf 86}, 3492 (2001) [arXiv:hep-ph/0012039].

\bibitem{Amore:2001uf}
P.~Amore, M.~C.~Birse, J.~A.~McGovern and N.~R.~Walet,
Phys.\ Rev.\ D {\bf 65} (2002) 074005 [arXiv:hep-ph/0110267].

\bibitem{Alford:2002kj}
M.~Alford and K.~Rajagopal,
JHEP {\bf 0206} (2002) 031 [arXiv:hep-ph/0204001].

\bibitem{Hong:1998tn}
D.~K.~Hong,
Phys.\ Lett.\ B {\bf 473} (2000) 118 [arXiv:hep-ph/9812510];
D.~K.~Hong,
Nucl.\ Phys.\ B {\bf 582} (2000) 451 [arXiv:hep-ph/9905523].

\bibitem{Beane:2000ms}
S.~R.~Beane, P.~F.~Bedaque and M.~J.~Savage,
Phys.\ Lett.\ B {\bf 483} (2000) 131 [arXiv:hep-ph/0002209].
%
\bibitem{Casalbuoni:2000na}
R.~Casalbuoni, R.~Gatto and G.~Nardulli,
Phys.\ Lett.\ B {\bf 498} (2001) 179 [Erratum-ibid.\ B {\bf 517}
(2001) 483] [arXiv:hep-ph/0010321].

\bibitem{Nardulli:2002ma}
G.~Nardulli,
Riv.\ Nuovo Cim.\  {\bf 25N3} (2002) 1 [arXiv:hep-ph/0202037].
%
\bibitem{Casalbuoni:2002st}
R.~Casalbuoni, F.~De Fazio, R.~Gatto, G.~Nardulli and M.~Ruggieri,
Phys.\ Lett.\ B {\bf 547} (2002)  229 [arXiv:hep-ph/0209105].

\bibitem{Alford:1999xc}
M.~G.~Alford, J.~Berges and K.~Rajagopal,
Phys.\ Rev.\ Lett.\  {\bf 84} (2000) 598 [arXiv:hep-ph/9908235].

\bibitem{Son:1998uk}
D.~T.~Son,
Phys.\ Rev.\ D {\bf 59} (1999) 094019 [arXiv:hep-ph/9812287].

\bibitem{Klevansky:qe}
S.~P.~Klevansky,
Rev.\ Mod.\ Phys.\  {\bf 64} (1992) 649.
%
\bibitem{Steiner:2002gx}
A.~W.~Steiner, S.~Reddy and M.~Prakash,
Phys.\ Rev.\ D {\bf 66}, 094007 (2002) [arXiv:hep-ph/0205201].

\bibitem{russi}L.~P.~Gorkov, T.~K.~Melik-Barchudarov, Zh. Eksp. Teor. Fiz. {\bf 40} (1961)
1452.
\end{thebibliography}
\end{document}